\documentclass[11pt]{article}

\usepackage{times}
\usepackage{amssymb}
\usepackage{amsmath}
\usepackage{amsbsy}
\usepackage{amsfonts}
\usepackage{eucal}
\newtheorem{theo}{Theorem}[section]
\newtheorem{lm}[theo]{Lemma}
 
\newcommand{\ket}[1]{\left\vert #1 \right\rangle}
\newcommand{\bra}[1]{\left\langle #1 \right\vert}
\newcommand{\braket}[2]{\left\langle #1 \left\vert #2 \right\rangle\right.}
\newcommand{\average}[3]{\left\langle #1\left | #2 \right | #3 \right\rangle}
\newcommand{\project}[2]{\left\vert #1 \right\rangle\left\langle #2 \right\vert}

\newcommand{\av}[1]{\left\lvert #1 \right\rvert}
\newcommand{\innerprod}[2]{\left\langle #1 , #2 \right\rangle}

\newcommand{\zz}{\mathbb Z}
 
\newcommand{\cc}{\mathbb C} 
\newcommand{\ff}{\mathbb F} 
 
\newcommand{\pp}{\mathbb P}

\newcommand{\mm}{\mathbb M}
 
\newcommand{\vph}{\varphi}

\newcommand{\bsym}{\boldsymbol}

\newcommand{\Tr}{\mathrm{Tr}} 
\newcommand{\qed}{\mbox{\rule{1.6mm}{4.3mm}}}
\newcommand{\one}{\mbox{\large{$1 \hspace{-0.95mm}  {\bf l}$}}}

\begin{document}

\title{A new proof for the existence of mutually unbiased 
bases\thanks{This work was supported in part by the Defense Advanced Research
Projects Agency (DARPA) project MDA 972-99-1-0017 (note that the content of
this paper does not necessarily reflect the position or the policy of the
government
and no official endorsement should be inferred), and in part by the U.S. Army
Research Office/DARPA under contract/grant number DAAD 19-00-1-0172.
%\newline\mbox{\hspace{4mm}}
The research of Farrokh Vatan was partly performed at the Jet Propulsion 
Laboratory (JPL), California Institute of Technology, under contract with 
National Aeronautics and Space Administration (NASA). 
%The Revolutionary computing
%Technologies Program of the JPL's Center for Integrated Space Microsystems
%(CISM) supported his work.
%\newline\mbox{\hspace{4mm}}
Emails: som@ee.ucla.edu, boykin@ee.ucla.edu, vwani@ee.ucla.edu, 
        Farrokh.Vatan@jpl.nasa.gov}}
\author{Somshubhro Bandyopadhyay$^{1,2}$, P. Oscar Boykin$^1$, 
        Vwani Roychowdhury$^1$,
        Farrokh Vatan$^{1,3}$ \\
        {\small 1. Electrical Engineering Department,  
                   UCLA, Los Angeles, CA 90095} \\
        {\small 2. Department of Physics, Bose Institute, 
                   Calcutta- 700009, India} \\
        {\small 3. Jet Propulsion Laboratory, California Institute of 
                   Technology 4800 Oak Grove Drive Pasadena, CA 91109}} 
\date{ }

\maketitle

\begin{abstract}
We develop a strong connection between maximally commuting bases of
orthogonal unitary matrices and mutually unbiased bases. A necessary
condition of the existence of mutually unbiased bases for any finite
dimension is obtained. Then a constructive proof of the existence of
mutually unbiased bases for dimensions that are powers of  primes
is presented. It is also proved that in any dimension $d$ the number of mutually 
unbiased bases is at most $d+1$. An explicit representation of mutually
unbiased observables in terms of Pauli matrices are provided for $d=2^m$.  
\end{abstract}   

\section{Introduction}

A $d$--level quantum system is described by a density operator  $\rho$
that requires $d^{2}-1$ real numbers for its complete specification.  A maximal
orthogonal quantum test performed on such a system has, without degeneracy, $d$  
possible outcomes, providing $d-1$ independent probabilities. It follows  that in
principle one requires at least $d+1$ different orthogonal measurements for  
complete state determination. 
\iffalse Here one should also note that for complete state
determination the underlying assumption is the availability of a large number
of identically prepared copies of the quantum system whereby measurements are
performed on different subensembles. 
\fi

Since the quantum mechanical description of a physical system  is characterized
in terms of probabilities of outcomes of conceivable  experiments consistent
with quantum formalism, in order to obtain full information  about the system
under consideration we need to perform  measurements  on a large number of
identically prepared copies of the system. The different measurements
are performed on several subensembles. However, there  may be redundancy in the
measurement results as the probabilities will not, in general, be independent
of each other unless a  minimal set of  measurements satisfying appropriate
criteria is specified. This minimal set need not be necessarily optimal  in the
sense it may not serve the best way to ascertain the quantum state.  However,
intuitively speaking, a minimal set of  measurements can be reasonably  close
to an  optimal set if they mutually differ as much as possible, thereby ruling 
out possible overlaps in the results which become crucial in case of error
prone measurements.  The characterization and proving the existence of such a 
minimal set of  measurements  for complete quantum state determination is
therefore of  fundamental importance. 

It has been shown that measurements in a special class of bases, i.e.  mutually
unbiased bases, not only form a minimal set but also provide the optimal
way  of determining a quantum state. Mutually unbiased measurements (MUM),
loosely speaking, correspond to measurements that are as different as they can
be so that each  measurement gives as much new information as one can obtain
from the system  under consideration. In other words the MUM operators are 
maximally  noncommuting among themselves. If the result of one MUM can be
predicted with  certainty, then all possible outcomes of every other
measurement, unbiased to  the previous one are equally likely. 

As noted earlier mutually unbiased bases (MUB) have a special role in
determining the state of a finite dimensional quantum system. Ivanovic
\cite{ivanovic81} first introduced the concept of MUB  in the context of
quantum state determination, where he proved the existence of such bases when
the dimension is a prime by an explicit construction. Later Wootters and Fields
\cite{wootters89} showed that measurements in MUB provide the minimal
as well as optimal way of complete specification of the density matrix. The
optimality is understood in the sense of minimization of statistical errors in
the measurements.  By explicit construction they showed the existence of MUB
for prime power dimensions and proved that for any dimension $d$ there can be
at most $d+1$ MUB. However the existence of MUB for other composite
dimensions which are not power of a prime still remains an open problem. 

In this paper we give a constructive proof of the results earlier obtained
by Ivanovic,  Wootters, and Fields \cite{ivanovic81, wootters89} with a totally 
different method.  The two distinct features of our new proof are:    

\begin{itemize} 
 
\item 
Our approach is based on developing an interesting connection between maximal 
commuting bases of orthogonal unitary matrices and mutually unbiased bases,
whereby  we find a  necessary condition for existence of MUB in any dimension.
We then provide a constructive proof of existence of MUB in composite
dimensions which are power of a prime.  
This allows us to connect encryption of quantum bits \cite{boykin00}, 
which uses unitary bases of operators, 
to quantum key distribution, which uses mutually unbiased bases of quantum
systems.   

\item  
Another advantage of our method is that  we provide an explicit construction of
the MUB observables (operators) as tensor product of the Pauli matrices for
dimensions $d=2^m$. This answers a critical related question:  how can 
these mutually unbiased measurements be actually performed and 
what are the observables to which these measurements
correspond to. When $d=2$ the mutually unbiased operators are the three Pauli
matrices, but unfortunately this observation cannot be generalized in a
straightforward way to higher dimension. In addition to  the obvious importance of
mutually unbiased bases in the  context of quantum state determination and
foundations of quantum mechanics, recently it has also found useful
applications in quantum cryptography where it has been demonstrated that
using higher dimensional quantum systems for key distribution has possible
advantages over qubits, and mutually unbiased bases play a key role in such 
a key distribution scheme \cite{bechmann00a, bechmann00b}. 
Thus the  fact that we provide an explicit construction of
the MUB observables can turn out to be crucial in the application of  MUB in quantum
cryptography with systems with more than two states.   
\end{itemize}

Before continuing it is useful to provide a formal definition of mutually 
unbiased bases.
%Before we continue 
%any further, we present the formal definition of mutually unbiased bases.  

\vspace{4mm}
\noindent

{\bf Definition.} Let
$B_1=\left\{\ket{\varphi_1},\ldots,\ket{\varphi_d}\right\}$ and
$B_2=\left\{\ket{\psi_1},\ldots,\ket{\psi_d}\right\}$be two orthonormal
bases in the $d$ dimensional state space. They are said to be
{\bf mutually unbiased bases (MUB)}   
if  and only if
$\left\vert\braket{\vph_i}{\psi_j}\right\vert=\frac{1}{\sqrt{d}}$, for every
$i,j=1,\ldots,d$. A set $\left\{ {\cal B}_1,\ldots,{\cal B}_m\right\}$ of
orthonormal bases in $\cc^d$ is called a {\em set of mutually unbiased bases}
(a set of MUB) if each pair of bases ${\cal B}_i$ and ${\cal B}_j$ are
mutually unbiased.  \vspace{4mm}  

The simplest example of a complete set of MUB is obtained in the case of spin 1/2 particle where
each unbiased basis consists of the normalized eigenvectors of the three Pauli
matrices respectively. However, the analysis of a set of  MUB corresponding to a two level
quantum system does not capture one of the basic features of MUB, i.e., its 
importance in determining the quantum state. In the case of two level systems, the
density operator has three independent parameters and almost any choice of the
three measurements is sufficient to have the complete knowledge of the system.
This is not true in general for any other dimension greater than two, where the existence
of MUB becomes more crucial in the context of minimal number of required
measurements for quantum state determination.

In Section~\ref{section-2} we show the existence of $p+1$ MUB in the space
$\cc^p$, for any prime $p$. This result first shown by Ivanovic \cite{ivanovic81}
by explicitly defining the mutually unbiased bases. Here we show that these bases 
are in fact bases each consists of eigenvectors of the unitary operators
\[ Z,\, X,\, XZ, \ldots,\, XZ^{d-1} , \]  where $X$ and $Z$ are
generalizations of Pauli operators to the quantum systems with more than two
states (see, e.g., \cite{gottesman99, gottesman00}). 
 
In Section~\ref{section-3} we show that there is a useful connection between mutually 
unbiased bases and special types of bases for the space of the square matrices. These
bases consist of orthogonal unitary matrices which can be grouped in
maximal classes of commuting matrices. As a result of this connection we show
that every MUB over $\cc^d$ consists of at most $d+1$ bases.

Finally, in Section~\ref{section-4} we present our construction of MUB over $\cc^d$ 
when $d$ is a prime power. The basic idea of our construction is as follows.
When $d=p^{m}$,  imagine the system consists of $m$ subsystems each of
dimension $p$. Then the total number of measurements on the whole system,
viewed as performing measurement on every subsystem in their respective MUB
is $(p+1)^m$. We show that these $(p+1)^m$ operators fall into $p^m+1$ maximal
noncommuting classes where members of  each class commute among themselves.
The bases formed by eigenvectors of each such  mutually noncommuting class
are mutually unbiased.  It should be mentioned  that the operators in each
maximal commuting class have the same  structure as the {\bf stabilizers} of
{\em additive quantum error correcting codes}
(see, e.g., \cite{calderbank97,calderbank98, gottesman99}).

One of the referees has brought to our attention that
there is a close connection between the MUB problem and the problem 
of determining arrangements of lines in the Grassmannian spaces so that they are as far apart 
as possible \cite{calderbank97b} (see also \cite{calderbank99}).
This problem (and some other combinatorial problems discussed in 
\cite{calderbank97b}) can be related to the problem of finding the maximum
number of lines through the origin of $\cc^d$ that are either perpendicular or 
are at angle $\theta$, where $\cos \theta=1/\sqrt{d}$. 
Any MUB $\cal M$ defines such a line--set: consider all lines through the 
origin defined by all vectors in the bases of $\cal M$.
In \cite{calderbank97b}, for the case of $d=2^m$, with an approach similar
to the one presented in this paper, such line--sets are constructed.

\vspace{4mm}
\noindent
{\bf Notation.}
Let $\mm_d(\cc)$ be the set of $d\times d$ complex matrices. In a natural way, 
the set $\mm_d(\cc)$ is a 
$d^2$--dimensional linear space. Each matrix $A$ in $\mm_d(\cc)$ can be also
naturally considered as a $d^2$--dimensional complex vector $\ket{v_A}$, 
where the entries
of the matrix $A$ being regarded as the components of the vector $\ket{v_A}$.
In this way, for matrices 
$A,B\in\mm_d(\cc)$ we can define the inner product $\innerprod{A}{B}$ of matrices
as the inner product $\braket{v_A}{v_B}$ of vectors. It is easy to check that
\[  \innerprod{A}{B}=\Tr\bigl(A^\dagger B\bigr).  \]
We say the matrices $A,B\in\mm_d(\cc)$ 
are {\bf orthogonal} if and only if $\innerprod{A}{B}=0$.

\section{Construction of sets of MUB for prime dimensions}
\label{section-2}

Ivanovic \cite{ivanovic81} for the first time showed that for any prime dimension
$d$, there is a set of $d+1$ mutually unbiased bases. In that paper the bases
are given explicitly. Here we show that there is a nice symmetrical structure
behind these bases, and their existence can be derived as a consequence of 
properties of Pauli operators on $d$--state quantum systems.
The core of our construction is the following theorem.

\begin{theo}
Let ${\cal B}_1=\{\,\ket{\vph_1},\ldots,\ket{\vph_d}\,\}$ be an orthonormal 
basis in $\cc^d$. Suppose that there is a unitary operator $V$ such that 
$V\ket{\vph_j}=\beta_j\ket{\vph_{j+1}}$, where $|\beta_j|=1$ and 
$\ket{\vph_{d+1}}=\ket{\vph_1}$; i.e., $V$ applies a {\em cyclic shift modulo
a phase}\/ on the elements of the basis ${\cal B}_1$. Assume that the orthonormal basis
${\cal B}_2=\{\,\ket{\psi_1},\ldots,\ket{\psi_d}\,\}$ consists of eigenvectors
of $V$. Then ${\cal B}_1$ and ${\cal B}_2$ are MUB.
\end{theo}

\noindent
{\bf Proof.}
Assume that $V\ket{\psi_k}=\lambda_k\ket{\psi_k}$. Then $|\lambda_k|=1$. 
Now, for every $k=1,\ldots,d$, we have
\begin{eqnarray*}
 \av{\braket{\psi_k}{\vph_1}} & = & \av{{\lambda_k}^*\average{\psi_k}{V}{\vph_1}} \\
 & = & \av{\beta_1\braket{\psi_k}{\vph_2}} \\ 
 & = & \av{\braket{\psi_k}{\vph_2}} .
\end{eqnarray*}
A similar argument shows
\[ \av{\braket{\psi_k}{\vph_1}}=\av{\braket{\psi_k}{\vph_2}}=\cdots=
     \av{\braket{\psi_k}{\vph_d}} .\]
Therefore,
\[ {\av{\braket{\psi_k}{\vph_j}}}^2=\frac{1}{d}, \qquad 1\leq j\leq d. \]
Thus ${\cal B}_1$ and ${\cal B}_2$ are MUB. \qed

\vspace{7mm}
Throughout this section, we suppose that $d$ is a prime number, and {\em all algebraic
operations are modulo $d$.} We consider $\{\, \ket{0},\ket{1},\ldots,\ket{d-1}\,\}$ as 
the standard basis of $\cc^d$. We define the unitary
operators $X_d$ and $Z_d$ over $\cc^d$, as a natural generalization of Pauli operators
$\sigma_x$ and $\sigma_z$:
\begin{eqnarray}
 X_d\ket{j} & = & \ket{j+1},  \label{x-d} \\
 Z_d\ket{j} & = & \omega^j\ket{j},  \label{z-d}
\end{eqnarray}
where $\omega$ is a $d^{\mathrm{\,th}}$ root of unity; more specifically 
$\omega=\exp(2\pi i/d)$. We are interested in unitary operators of the form
$X_d\left(Z_d\right)^k$. Note that
\[ X_d\left(Z_d\right)^k\ket{j} = \bigl(\omega^k\bigr)^j\ket{j+1} .\]

\begin{theo}
For $0\leq k,\ell\leq d-1$, the eigenvectors of $X_d\left(Z_d\right)^k$ are cyclically
shifted under the action of $X_d\left(Z_d\right)^{\ell}$.
\end{theo}

\noindent
{\bf Proof.}
The eigenvectors of $X_d\left(Z_d\right)^k$ are 
\begin{equation}   
  \ket{\psi_t^k}=\frac{1}{\sqrt{d}}\sum_{j=0}^{d-1}\left(\omega^t\right)^{d-j}
      \big(\omega^{-k}\big)^{s_j}\ket{j}, \qquad t=0,\ldots,d-1, 
\label{eigenvec} 
\end{equation}
where $s_j=j+\cdots+(d-1)$. Then $\ket{\psi_t^k}$ is an eigenvector of 
$X_d\left(Z_d\right)^k$ with eigenvalue $\omega^t$, because
\begin{eqnarray*}
 X_d\left(Z_d\right)^k\ket{\psi_t^k} & = & \frac{1}{\sqrt{d}}
    \sum_{j=0}^{d-1}\big(\omega^t\big)^{d-j}\big(\omega^{-k}\big)^{s_j}
    \big(\omega^k\big)^j\ket{j+1} \\
  & = & \frac{1}{\sqrt{d}}
    \sum_{j=0}^{d-1}\big(\omega^t\big)^{d-j}\big(\omega^{-k}\big)^{s_{j+1}}
    \ket{j+1} \\ 
  & = & \frac{1}{\sqrt{d}}
    \sum_{j=0}^{d-1}\big(\omega^t\big)^{d-j+1}\big(\omega^{-k}\big)^{s_{j}}
    \ket{j} \\
  & = & \omega^t \ket{\psi_t^k}.   
\end{eqnarray*}
The action of $X_d\left(Z_d\right)^\ell$ on $\ket{\psi_t^k}$ is as follows:
\begin{eqnarray*}
 X_d\left(Z_d\right)^\ell\ket{\psi_t^k} & = & \frac{1}{\sqrt{d}}
    \sum_{j=0}^{d-1}\big(\omega^t\big)^{d-j}\big(\omega^{-k}\big)^{s_j}
    \big(\omega^\ell\big)^j\ket{j+1} \\
  & = & \frac{1}{\sqrt{d}}
    \sum_{j=0}^{d-1}\big(\omega^t\big)^{d-j+1}\big(\omega^{-k}\big)^{s_{j-1}}
    \big(\omega^\ell\big)^{j-1}\ket{j} \\ 
  & = & \frac{\omega^{t-\ell}}{\sqrt{d}}
    \sum_{j=0}^{d-1}\big(\omega^t\big)^{d-j}\big(\omega^{-k}\big)^{s_{j}}
    \big(\omega^{-k}\big)^{j-1}\big(\omega^\ell\big)^{j}\ket{j} \\
  & = & \frac{\omega^{t+k-\ell}}{\sqrt{d}}
    \sum_{j=0}^{d-1}\big(\omega^t\big)^{d-j}\big(\omega^{-k}\big)^{s_{j}}
    \big(\omega^{\ell-k}\big)^j\ket{j} \\
  & = & \frac{\omega^{t+k-\ell}}{\sqrt{d}}
    \sum_{j=0}^{d-1}\big(\omega^{t+k-\ell}\big)^{d-j}
    \big(\omega^{-k}\big)^{s_{j}}\ket{j} \\
  & = & \omega^{t+k-\ell}\ket{\psi^k_{t+k-\ell}} . \quad\qed
\end{eqnarray*}

Note that the standard basis $\{\, \ket{0},\ket{1},\ldots,\ket{d-1}\,\}$ is the 
set of the eigenvectors of $Z_d$. From (\ref{eigenvec}) it follows that the
$\av{\braket{j}{\psi^k_t}}^2=\frac{1}{d}$. Therefore, we have proved the following
construction.

\begin{theo}
For any prime $d$, the set of the bases each consisting of the eigenvectors of 
\[ Z_d,\; X_d,\; X_dZ_d,\; X_d\left(Z_d\right)^2,\ldots,\; X_d\left(Z_d\right)^{d-1}, \]
form a set of $d+1$ mutually unbiased bases.
\label{prime}
\end{theo}

\vspace{4mm}
\noindent
{\bf Example $d=2$.}
By Theorem~\ref{prime}, the eigenvectors of the operators $\sigma_z$, $\sigma_x$, and
$\sigma_x\,\sigma_z$ form a set of mutually unbiased bases; i.e., the following set
\[ \begin{array}{l}
  \left\{ \ket{0}, \ket{1} \right\}, \\
  \left\{ \frac{\ket{0}+\ket{1}}{\sqrt{2}}, \frac{\ket{0}-\ket{1}}{\sqrt{2}}\right\}, \\
  \left\{ \frac{\ket{0}+i\ket{1}}{\sqrt{2}}, \frac{\ket{0}-i\ket{1}}{\sqrt{2}}\right\} .
\end{array} \]

\vspace{4mm}
\noindent
{\bf Example $d=3$.} The set of the eigenvectors of the following unitary matrices
form a set of MUB (here $\omega=\exp(2\pi i/3)$):
\[  \begin{pmatrix} 1 & 0 & 0 \\ 0 & 1 & 0 \\ 0 & 0 & 1 \end{pmatrix}, \qquad
    \begin{pmatrix} 0 & 0 & 1 \\ 1 & 0 & 0 \\ 0 & 1 & 0 \end{pmatrix}, \qquad
    \begin{pmatrix} 0 & 0 & \omega^2 \\ 1 & 0 & 0 \\ 0 & \omega & 0 \end{pmatrix}, \qquad
    \begin{pmatrix} 0 & 0 & \omega \\ 1 & 0 & 0 \\ 0 & \omega^2 & 0 \end{pmatrix}. 
   \]

\section{Bases for unitary operators and MUB}
\label{section-3}

In this section we study the close relation between MUB and a special type of bases
for $\mm_d(\cc)$. Here we are dealing with classes of commuting unitary matrices.
The following lemma shows that the maximum size of such class is $d$.

\begin{lm}
There are at most $d$ pairwise orthogonal commuting unitary matrices in $\mm_d(\cc)$.
\label{maximal-lm}
\end{lm}

{\bf Proof.}
Let $A_1,\ldots,A_m$ be pairwise orthogonal commuting unitary matrices in
$\mm_d(\cc)$. Then there is a unitary matrix $U$ such that the matrices 
$B_1,\ldots,B_m$, where $B_j=U\,A_j\,U^\dagger$, are diagonal. Moreover,
$\innerprod{B_j}{B_k}=\innerprod{A_j}{A_k}$; so $B_j$ and $B_k$ are orthogonal for 
$j\neq k$. Let $\ket{\bsym{b}_j}\in\cc^d$ be the diagonal of $B_j$.
Then $\innerprod{B_j}{B_k}=\braket{\bsym{b}_j}{\bsym{b}_k}$.
So the vectors $\ket{\bsym{b}_1},\ldots,\ket{\bsym{b}_m}$ are mutually orthogonal;
therefore, $m\leq d$.~\qed

\vspace{7mm}
Let ${\cal B}=\left\{\,U_1,U_2,\ldots,U_{d^2}\,\right\}$ be a {\em basis of unitary} 
matrices for $\mm_d(\cc)$. Without loss of generality, we can assume that $U_1=\one_d$, 
the identity matrix of order $d$. We say that the basis $\cal B$ is a 
{\bf maximal commuting basis} for $\mm_d(\cc)$ if $\cal B$ can be partitioned as
\begin{equation}
   {\cal B}=\big\{\one_d\big\}\bigcup {\cal C}_1\bigcup \cdots \bigcup {\cal C}_{d+1}, 
\label{mcb}
\end{equation}  
where each class ${\cal C}_j$ contains exactly $d-1$ commuting matrix from $\cal B$.
Note that $\{\one_d\}\bigcup {\cal C}_j$ is a set of $d$ commuting orthogonal unitary 
matrices, which by Lemma~\ref{maximal-lm} is maximal.

\begin{theo}
If there is a maximal commuting basis of orthogonal unitary matrices in 
$\mm_d(\cc)$, then there is a set of $d+1$ mutually unbiased bases.
\label{max-basis-theo}
\end{theo}

{\bf Proof.}
Let $\cal B$ be a maximal commuting basis of orthogonal unitary matrices in 
$\mm_d(\cc)$, where (\ref{mcb}) provides the decomposition of $\cal B$ into
maximal classes of commuting matrices. For any $1\leq j\leq d+1$, let
\[ {\cal C}_j=\left\{\, U_{j,1},U_{j,2},\ldots,U_{j,d-1} \,\right\}.  \]
%Without loss of generality, we can assume that all these matrices are normalized; i.e.,
%$\innerprod{U_{j,k}}{U_{j,k}}=1$.
We also define $U_{j,0}=\one_d$; then 
\[ {\cal C}_j'=\left\{\, U_{j,0},U_{j,1},U_{j,2},\ldots,U_{j,d-1} \,\right\} \]
is a maximal set of commuting orthogonal unitary matrices. Thus for each 
$1\leq j\leq d+1$, there is an orthonormal basis 
\[ {\cal T}_j=\left\{\, \ket {\psi^j_1},\ket{\psi^j_2},\ldots,\ket{\psi^j_d} \,\right\} \] 
such that every matrix $U_{j,t}$ (for $0 \leq t \leq d-1$) relative to the basis 
${\cal T}_j$ is diagonal. Let
\begin{equation}
  U_{j,t} = \sum_{k=1}^d \lambda_{j,t,k}\project{\psi^j_k}{\psi^j_k} . 
\label{diag-equ}
\end{equation}  
Let $M_j$ be a $d\times d$ matrix whose $k^{\mathrm{\,th}}$ row is the diagonal of the
right-hand side matrix of (\ref{diag-equ}); i.e.,
\[ M_j=\begin{pmatrix}
         \lambda_{j,0,1} & \lambda_{j,0,2} & \ldots & \lambda_{j,0,d} \\
         \lambda_{j,1,1} & \lambda_{j,1,2} & \ldots & \lambda_{j,1,d} \\
         \vdots          & \vdots          & \ddots & \vdots \\  
         \lambda_{j,d-1,1} & \lambda_{j,d-1,2} & \ldots & \lambda_{j,d-1,d}
        \end{pmatrix} . \]
Then $M_j$ is a {\em unitary} matrix. Note that the first row of $M_j$ is the 
constant vector $(1,1,\ldots,1)$. 
We consider the classes ${\cal C}_1$ and ${\cal C}_2$.
Then for $0\leq s,t\leq d-1$, the orthogonality condition implies
\[ \Tr\left( {U_{1,s}}^\dagger \; U_{2,t} \right) = d\,\delta_{s,0} \; \delta_{t,0}. \]
But, since 
$\Tr\left(\project{\psi^1_k}{\psi^2_\ell}\right)=\braket{\psi^1_k}{\psi^2_\ell}^*$,
\begin{eqnarray*}
 \Tr\left( {U_{1,s}}^\dagger \; U_{2,t} \right) & = & 
  \Tr\left( \sum_{k=1}^d\sum_{\ell=1}^d {\lambda_{1,s,k}}^*\lambda_{2,t,\ell}
    \ket{\psi^1_k}\braket{\psi^1_k}{\psi^2_\ell}\bra{\psi^2_\ell} \right) \\
 & = & \sum_{k=1}^d\sum_{\ell=1}^d {\lambda_{1,s,k}}^*\lambda_{2,t,\ell}
    \braket{\psi^1_k}{\psi^2_\ell}\Tr\left(\project{\psi^1_k}{\psi^2_\ell}\right) \\
 & = & \sum_{k=1}^d\sum_{\ell=1}^d {\lambda_{1,s,k}}^*\lambda_{2,t,\ell}
    \av{\braket{\psi^1_k}{\psi^2_\ell}}^2 .
\end{eqnarray*}
Therefore
\begin{equation}
  \sum_{k=1}^d\sum_{\ell=1}^d {\lambda_{1,s,k}}^*\lambda_{2,t,\ell}
    \av{\braket{\psi^1_k}{\psi^2_\ell}}^2 = d\,\delta_{s,0}\delta_{t,0},
    \qquad 0\leq s,t \leq d-1 .
\label{tr-equ}
\end{equation}
The system of equations (\ref{tr-equ}) can be written in the following matrix form
\[ A \, P = \Lambda,  \]
where 
\begin{eqnarray*}
 A & = & {M_1}^* \otimes M_2, \\
 P & = & \left(\av{\braket{\psi^1_1}{\psi^2_1}}^2,
         \av{\braket{\psi^1_1}{\psi^2_2}}^2, \ldots,
         \av{\braket{\psi^1_d}{\psi^2_d}}^2 \right)^{\mathrm T}, \\
 \Lambda & = & (d,0,0,\ldots,0)^{\mathrm T}.
\end{eqnarray*}
Note that $A$ is a unitary matrix and its first row is the constant vector
$(1,1,\ldots,1)$. Then from $P=A^{-1}\Lambda$ it follows
\[ \textstyle \av{\braket{\psi^1_s}{\psi^2_t}}^2=\frac{1}{d},\qquad 1\leq s,t\leq d .\]
By repeating the same argument for the classes ${\cal C}_j$ and ${\cal C}_k$, 
we conclude that
\[ \left\{\, {\cal T}_1,\ldots,{\cal T}_{d+1}\,\right\} \]
is a set of MUB. \qed

\vspace{7mm}
Before we continue, we prove the following useful simple lemma.

\begin{lm}
For any integers $m$ and $n$ such that $0 < m \leq n$ we have
\[ \sum_{k=1}^n e^{2\pi i\frac{mk}{n}} = 0.  \]
\label{number-theo-lemma}
\end{lm}

\noindent
{\bf Proof.} We have
\[ \sum_{k=1}^n \left(e^{2\pi i\frac{m}{n}}\right)^k = e^{2\pi i\frac{m}{n}}
  \frac{\left(e^{2\pi i\frac{m}{n}}\right)^n-1}{e^{2\pi i\frac{m}{n}}-1} = 0.\quad\qed \]

The converse of Theorem~\ref{max-basis-theo}, in the following sense, holds.

\begin{theo}
Let ${\cal B}_1,\ldots,{\cal B}_m$ be a set of MUB in $\cc^d$. Then there are $m$
classes ${\cal C}_1,\ldots,{\cal C}_m$ each consisting of $d$ commuting unitary 
matrices such that matrices in ${\cal C}_1\bigcup\cdots\bigcup{\cal C}_m$ are pairwise
orthogonal.
\label{mub-bases}
\end{theo}

\noindent
{\bf Proof.} 
Suppose that
\[ {\cal B}_j=\left\{\, \ket{\psi^j_1},\ldots,\ket{\psi^j_d}\,\right\}. \]
Then
\[ \braket{\psi^j_s}{\psi^j_t} = \delta_{s,t}, \qquad 1\leq s,t\leq d,\]
and
\[  \av{\braket{\psi^j_s}{\psi^k_t}}^2 = \frac{1}{d}, 
        \qquad 1\leq j<k\leq d,\ 1\leq s,t\leq d.\]
We label the matrices in the class ${\cal C}_j$ as 
\[ {\cal C}_j = \left\{\, U_{j,0},U_{j,1},\ldots,U_{j,d-1}\,\right\} ,\]
where
\[ U_{j,t} = \sum_{k=1}^d e^{2\pi i\frac{tk}{d}} 
             \project{\psi^j_k}{\psi^j_k} , \qquad 0\leq t \leq d-1 . \]
Note that $U_{j,0}=\one_d$. Then $U_{j,s}$ and $U_{j,t}$ are commuting, because
both are diagonal relative to the basis ${\cal B}_j$.
We now show that all these matrices are orthogonal. First we note that
\begin{eqnarray*}
 \innerprod{U_{j,s}}{U_{k,t}} & = & \Tr\left( {U_{j,s}}^\dagger \; U_{k,t} \right) \\
 & = & \sum_{x=1}^d\sum_{y=1}^d e^{2\pi i \frac{ty-sx}{d}} \Tr\left(
       \ket{\psi^j_x\vphantom{\psi^k_y}}\braket{\psi^j_x}{\psi^k_y}\bra{\psi^k_y} \right)\\
 & = & \sum_{x=1}^d\sum_{y=1}^d e^{2\pi i \frac{ty-sx}{d}}
       \av{\braket{\psi^j_x}{\psi^k_y}}^2 .
\end{eqnarray*}
Thus, by Lemma~\ref{number-theo-lemma}, if $j = k$, then
\begin{eqnarray*}
 \innerprod{U_{j,s}}{U_{j,t}} & = & 
       \sum_{x=1}^d\sum_{y=1}^d e^{2\pi i \frac{ty-sx}{d}} \delta_{x,y} \\
 & = & \sum_{x=1}^d e^{2\pi i \frac{x(t-s)}{d}} \\
 & = & d\,\delta_{s,t} .
\end{eqnarray*}
If $j \neq k$ and $(s,t) \neq (0,0)$, then
\begin{eqnarray*}
 \innerprod{U_{j,s}}{U_{k,t}} & = &
    \sum_{x=1}^d\sum_{y=1}^d e^{2\pi i \frac{ty-sx}{d}} \frac{1}{d} \\
 & = & \frac{1}{d}
       \left( \sum_{x=1}^d e^{2\pi i \frac{sx}{d}} \right) ^*
       \left( \sum_{y=1}^d e^{2\pi i \frac{ty}{d}} \right) \\
 & = & 0 .\ \qed
\end{eqnarray*}

As an immediate corollary of the above theorem, we have the following upper bound
on the size of a set of MUB.

\begin{theo}
Any set of mutually unbiased bases in $\cc^d$ contains at most $d+1$ bases.
\end{theo}

\noindent
{\bf Proof.}
If a set of MUB contains $m$ bases, then by Theorem \ref{mub-bases}, there are 
at least $1+m(d-1)$ pairwise orthogonal matrices in the $d^2$--dimensional space
$\mm_d(\cc)$. Therefore, $1+m(d-1) \leq d^2$, thus $m\leq d+1$.~\qed

\section{Construction of a set of MUB for prime powers}
\label{section-4}

\subsection{The Pauli group}

To construct a maximal set of MUB in ${\cal H}=\cc^{p^m}$, where $p$ is a prime number,
we consider the Hilbert space $\cal H$ as tensor product of $m$ copies of
$\cc^p$; i.e., 
\[  {\cal H}=\underbrace{\cc^p\otimes\cdots\otimes\cc^p}_{m\ \mathrm{times}} .  \]
Like the case of $\cc^p$, we build a set of MUB as the sets of eigenvectors of
special types of unitary operators on the background space $\cal H$. On the 
space $\cc^p$ we considered the generalized Pauli operators $X_p$ and $Z_p$,
defined by equations (\ref{x-d}) and (\ref{z-d}). On the space $\cal H$, 
we consider the tensor products of operators $X_p$ and $Z_p$. 
%For simplicity, we denote these operators by $X$ and $Z$, when the
%value of the prime $p$ is clear. We also define $Y=X\,Z$.

We denote the finite field $\{0,1,\ldots,p-1\}$ by $\ff_p$.
Let $\omega=e^{2\pi i/d}$ be a primitive $p^{\mathrm{th}}$ root of unity. Then
\[ Z_p \, X_p = \omega \, X_p \, Z_p .\]
Therefore, if $U_1=\left(X_p\right)^{k_1}\left(Z_p\right)^{\ell_1}$ and
$U_2=\left(X_p\right)^{k_2}\left(Z_p\right)^{\ell_2}$ then
\begin{equation}
 U_2 \, U_1 = \omega^{k_1\ell_2-k_2\ell_1} U_1 \, U_2 .
\label{u2-u1}
\end{equation}
We are interested on unitary operators on ${\cal H}=\cc^p\otimes\cdots\otimes\cc^p$
(the tensor product of $m$ copies of $\cc^p$) of the form
\begin{equation}
  U = M_1 \otimes \cdots \otimes M_m , \qquad
      \mbox{where $M_j=\left(X_p\right)^{k_j} \left(Z_p\right)^{\ell_j}$, 
      $0\leq k_j,\ell_j \leq p-1$.}
\label{operator}
\end{equation}
To describe an operator of the form (\ref{operator}) it is enough to specify the
powers $k_j$ and $\ell_j$. So we represent an operator (\ref{operator}) by the
following vector of length $2m$ over the field $\ff_p$:
\[  (k_1,\ldots,k_m \,\vert\, \ell_1,\ldots,\ell_m) ,\]
or equivalently as 
\[ X_p(k_1,\ldots,k_m) \, Z_p(\ell_1,\ldots,\ell_m) . \]
If we let $\alpha=(k_1,\ldots,k_m)$ and $\beta=(\ell_1,\ldots,\ell_m)$, then
$\alpha,\beta\in{\ff_p}^m$ and we denote the corresponding operator by
\[ X_p(\alpha)\,Z_p(\beta) . \]

The {\bf Pauli group} $\pp(p,m)$ is the group of all unitary operators on
${\cal H}=\cc^p\otimes\cdots\otimes\cc^p$ (the tensor product of $m$ copies of $\cc^p$)
of the form
\begin{equation}
  \omega^j \, X_p(\alpha) \, Z_p(\beta), 
\label{pauli}
\end{equation}
for some integer $j \geq 0$ and vectors $\alpha,\beta\in{\ff_p}^m$, where 
$\omega=\exp(2\pi i/p)$. In this section we are mainly interested in the subset
$\pp_0(p,m)$ of $\pp(p,m)$ of the operators of the form (\ref{pauli}) with $j=0$.
Note that $\pp_0(p,m)$ is not a subgroup, but generators of subgroups of 
the Pauli group can always be considered as subsets of $\pp_0(p,m)$.

If the operators $U$ and $U'$ in $\pp_0(p,m)$ are represented by the vectors 
\[ (k_1,\ldots,k_m \,\vert\, \ell_1,\ldots,\ell_m) \qquad \mbox{and} \qquad 
   (k_1',\ldots,k_m' \,\vert\, \ell_1',\ldots,\ell_m'), \] 
respectively, then $U$ and $U'$ are commuting if and only if
\[  \sum_{j=1}^m k_j\ell_j' -\sum_{j=1}^m k_j'\ell_j = 0 \mod p .\]
We can state this condition equivalently in the following form.

\begin{lm}
If $U=X_p(\alpha)\,Z_p(\beta)$ and $U'=X_p(\alpha')\,Z_p(\beta')$, 
for $\alpha,\beta,\alpha',\beta'\in{\ff_p}^m$, then $U$ and $U'$ are commuting
if and only if
\begin{equation}
  \alpha\cdot\beta'-\alpha'\cdot\beta=0 \mod p. 
\label{commuting-equ}
\end{equation}
\end{lm}

A set $X_p(\alpha_1)\,Z_p(\beta_1),\ldots,X_p(\alpha_t)\,Z_p(\beta_t)$ of operators in
$\pp_0(p,m)$ is represented by the $t\times(2m)$ matrix
\[ \left ( \begin{array}{c|c}
   \alpha_1 & \beta_1 \\ \vdots & \vdots \\ \alpha_t & \beta_t
   \end{array} \right ) . \]

Before we continue, we would like to get an explicit formula for the action of a
$\pp_0(p,m)$ operator $X_p(\alpha)\,Z_p(\beta)$. Let $\alpha=(\alpha_1,\ldots,\alpha_m)$ 
and $\beta=(\beta_1,\ldots,\beta_m)$. The standard basis of the Hilbert space
${\cal H}=\cc^p\otimes\cdots\otimes\cc^p$ consists of the vectors $\ket{j_1\cdots j_m}$,
where $(j_1,\ldots,j_m)\in{\ff_p}^m$. Then
\[ X_p(\alpha)\,Z_p(\beta) \ket{j_1\cdots j_m} =
    \omega^{j_1\beta_1+\cdots+j_m\beta_m}\ket{(j_1+\alpha_1)\,\cdots\,(j_m+\alpha_m)} . \]
Equivalently,
\begin{eqnarray}
  X_p(\alpha)\,Z_p(\beta)\ket{a} & = & \omega^{a\cdot\beta}\ket{a+\alpha},
                                       \qquad a\in{\ff_p}^m, \\ \label{pauli-action-1}
  X_p(\alpha)\,Z_p(\beta) & = & \sum_{a\in{\ff_p}^m} \omega^{a\cdot\beta}
                                \project{a+\alpha}{a}, \label{pauli-action-2}
\end{eqnarray}
where the operations are in the field $\ff_p$.

\begin{theo}
Let $U=X_p(\alpha)\,Z_p(\beta)$ and $U'=X_p(\alpha')\,Z_p(\beta')$ be operators in
$\pp_0(p,m)$. If $U \neq U'$, i.e., $(\alpha,\beta)\neq(\alpha',\beta')$, 
then the operators $U$ and $U'$ are orthogonal.
\end{theo}

\noindent
{\bf Proof.}
We have 
\begin{eqnarray*}
 \innerprod{U}{U'} & = & \Tr\left( U^\dagger \; U' \right) \\
 & = & \Tr\left(\sum_{a\in{\ff_p}^m}\sum_{b\in{\ff_p}^m} 
       \omega^{\beta'\cdot b-\beta\cdot a} 
       \ket{a}\braket{a+\alpha}{b+\alpha'}\bra{b}  \right) \\
 & = & \sum_{a\in{\ff_p}^m} \omega^{\beta'\cdot b-\beta\cdot a}
       \braket{a+\alpha}{a+\alpha'} .
\end{eqnarray*}
If $\alpha\neq\alpha'$, then $\braket{a+\alpha}{a+\alpha'}=0$, for every $a\in{\ff_p}^m$.
Thus in this case $\innerprod{U}{U'}=0$. If $\alpha=\alpha'$ and $\beta\neq\beta'$
then, by Lemma~\ref{number-theo-lemma},
\begin{eqnarray*}
 \innerprod{U}{U'} & = & \sum_{a\in{\ff_p}^m} \omega^{(\beta'-\beta)\cdot a} \\
    & = & 0.\quad \qed
\end{eqnarray*}

\subsection{The general construction}

Our scheme for constructing a set of MUB is based on Theorem~\ref{max-basis-theo}.
The maximal commuting orthogonal basis for $\mm_{p^m}(\cc)$ with partition of the form
(\ref{mcb}) is such that each class $\{\one_p\}\bigcup{\cal C}_j$, in the following sense,
is a {\em linear}\/ space of operators in the Pauli group $\pp(p,m)$. Let 
\[ X_p(\alpha_1)\,Z_p(\beta_1),\ldots,X_p(\alpha_{p^m})\,Z_p(\beta_{p^m}) \]
be the operators in the class $\{\one_p\}\bigcup{\cal C}_j$. We say that this class is
{\bf linear} if the set of the vectors 
\[ {\cal E}_j=\{\, (\alpha_1\vert\beta_1),\ldots,(\alpha_{p^m}\vert\beta_{p^m}) \,\} \] 
form an $m$--dimensional subspace of ${\ff_p}^{2m}$. In this case, to specify a linear 
class, it is enough to present a basis for the subspace ${\cal E}_j$.
Such a basis can be represented by an $m\times (2m)$ matrix. So instead of listing all
operators in the classes ${\cal C}_1,\ldots,{\cal C}_{p^m+1}$, we could simply list the 
$p^m+1$ matrices representing the bases of these classes. 

More specifically, the bases of linear classes of operators in our construction are 
represented by the matrices
\[ (0_m\vert\one_m),\quad (\one_m\vert A_1),\quad \ldots,\quad (\one_m\vert A_{p^m}) , \] 
where $0_m$ is the all--zero matrix of order $m$
and each $A_j$ is an $m\times m$ matrix over $\ff_p$.
It easy to see what conditions should be imposed on the matrices $A_j$ so that the
requirements of Theorem~\ref{max-basis-theo} satisfied. The following lemma gives a simple
necessary and sufficient condition for operators in each class commuting.
Note that in a linear class of operators, if the basic operators are commuting then
any pair of operators in these class will commute.

\begin{lm}
Let $S$ be a set of $m$ operators in $\pp_0(p,m)$, and $S$ be represented by the matrix
$(\one_m \vert A)$, where $\one_m$ is the identity matrix of order $m$ and $A$ is 
an $m \times m$ matrix over $\ff_p$. Then the operators in $S$ are pairwise commuting
if and only if $A$ is a symmetric matrix.
\label{commuting-lemma}
\end{lm}

\noindent
{\bf Proof.}
Let $A=(a_{jk})$. Then, by (\ref{commuting-equ}), $S$ is a set of commuting operators
if and only if $a_{jk}-a_{kj}=0 \mod p$, for every $1\leq j<k \leq m$.
Since $a_{jk}\in\ff_p$, 
$S$ is a set of commuting operators if and only if $A$ is symmetric.~\qed

\vspace{7mm}
The other condition is that the classes ${\cal C}_j$ and ${\cal C}_k$ should be disjoint.
This condition is met if the span of the  matrices $(\one_m\vert A_j)$ and 
$(\one_m\vert A_k)$ are disjoint. The last condition is equivalent to
$\bsym{x} A_j\neq\bsym{x}A_k$, for every non--zero $\bsym{x}\in{\ff_p}^m$. 
The last condition is equivalent to $\det(A_j-A_k) \neq 0$.
Thus we can summarize our construction in the following theorem.

\begin{theo}
Let $\{A_1,\ldots,A_\ell\}$ be a set of symmetric $m\times m$ matrices over $\ff_p$
such that $\det(A_j-A_k) \neq 0$, for every $1\leq j<k\leq \ell$. 
Then there is a set of $\ell+1$ mutually unbiased bases on $\cc^{p^m}$.
\label{construction-theo}
\end{theo}

More specifically, the $\ell+1$ bases of the above theorem are represented by the matrices
\[ (0_m\vert\one_m),\quad (\one_m\vert A_1),\quad \ldots,\quad (\one_m\vert A_\ell). \]

\vspace{4mm}
\noindent
{\bf Example $d=4$.}
The four matrices (over $\ff_2=\{0,1\}$) which satisfy the conditions of 
Theorem~\ref{construction-theo} are
\begin{equation}
    \begin{pmatrix} 0 & 0 \\ 0 & 0 \end{pmatrix}, \quad
    \begin{pmatrix} 1 & 0 \\ 0 & 1 \end{pmatrix}, \quad
    \begin{pmatrix} 0 & 1 \\ 1 & 1 \end{pmatrix}, \quad
    \begin{pmatrix} 1 & 1 \\ 1 & 0 \end{pmatrix}. 
\label{d=4-example}
\end{equation}    
Therefore the classes of maximal commuting operators are
\begin{eqnarray*}
 {\cal C}_0 & = & \left\{\, Z\otimes I,\, I\otimes Z,\, Z\otimes Z \,\right\}, \\
 {\cal C}_1 & = & \left\{\, X\otimes I,\, I\otimes X,\, X\otimes X \,\right\}, \\
 {\cal C}_2 & = & \left\{\, Y\otimes I,\, I\otimes Y,\, Y\otimes Y \,\right\}, \\
 {\cal C}_3 & = & \left\{\, X\otimes Z,\, Z\otimes Y,\, Y\otimes X \,\right\}, \\
 {\cal C}_4 & = & \left\{\, Y\otimes Z,\, Z\otimes X,\, X\otimes Y \,\right\},
\end{eqnarray*}
where
\[ I=\begin{pmatrix} 1 & 0 \\ 0 & 1 \end{pmatrix}, \qquad
   X=\begin{pmatrix} 0 & 1 \\ 1 & 0 \end{pmatrix}, \qquad
   Y=\begin{pmatrix} 0 & -1 \\ 1 & 0 \end{pmatrix}=XZ, \qquad
   Z=\begin{pmatrix} 1 & 0 \\ 0 & -1 \end{pmatrix}. \]
%  
%  
%The corresponding bases are 
%\begin{eqnarray*}
% {\cal B}_0 & = & \left\{\, \ket{00},\ \ket{01},\ \ket{10},\ \ket{11} \,\right\} \\
% {\cal B}_1 & = & \ts\left\{\, \frac{1}{2}(\ket{00}+\ket{01}+\ket{10}+\ket{11}), \quad
%                  \frac{1}{2}(\ket{00}-\ket{01}-\ket{10}+\ket{11}), \right. \\
% & &       \ts \left . \frac{1}{2}(\ket{00}+\ket{01}-\ket{10}-\ket{11}), \quad
%                    \frac{1}{2}(\ket{00}-\ket{01}+\ket{10}-\ket{11}) \,\right\} \\
% {\cal B}_2 & = & \ts\left\{\, \frac{1}{2}(\ket{00}+i\ket{01}+i\ket{10}-\ket{11}), \quad
%                  \frac{1}{2}(\ket{00}-i\ket{01}-i\ket{10}+\ket{11}), \right. \\
% & &       \ts \left . \frac{1}{2}(\ket{00}+i\ket{01}-i\ket{10}+\ket{11}), \quad
%                    \frac{1}{2}(\ket{00}-i\ket{01}+i\ket{10}+\ket{11}) \,\right\} \\
% {\cal B}_3 & = & \ts\left\{\, \frac{1}{2}(\ket{00}+\ket{01}-i\ket{10}+i\ket{11}), \quad
%                  \frac{1}{2}(\ket{00}-\ket{01}+i\ket{10}+i\ket{11}), \right. \\
% & &       \ts \left . \frac{1}{2}(\ket{00}+\ket{01}+i\ket{10}-i\ket{11}), \quad
%                    \frac{1}{2}(\ket{00}-\ket{01}-i\ket{10}-i\ket{11}) \,\right\} \\
% {\cal B}_4 & = & \ts\left\{\, \frac{1}{2}(\ket{00}-i\ket{01}+\ket{10}+i\ket{11}), \quad
%                  \frac{1}{2}(\ket{00}+i\ket{01}-\ket{10}+i\ket{11}), \right. \\
% & &       \ts \left . \frac{1}{2}(\ket{00}+i\ket{01}+\ket{10}-i\ket{11}), \quad
%                    \frac{1}{2}(\ket{00}-i\ket{01}-\ket{10}-i\ket{11}) \,\right\} .
%\end{eqnarray*}  
%
%
We represent this basis explicitly. To this end, we naturally represent each
basis by a $4\times4$ matrix such that the $j^{\mathrm{\,th}}$ row of this matrix 
is the components of the $j^{\mathrm{\,th}}$ vector of the corresponding basis
with respect to the standard basis $\ket{00},\ket{01},\ket{10},\ket{11}$:
the first matrix is ${\cal B}_0=\one_4$, and
\begin{xalignat*}{3}
 {\cal B}_1=\frac{1}{2}\left(\begin{array}{rrrr} 
 1 & 1 & 1 & 1 \\ 1 & -1 & -1 & 1 \\ 1 & 1 & -1 & -1 \\ 1 & -1 & 1 & -1 
 \end{array}\right), &&
 {\cal B}_2=\frac{1}{2}\left(\begin{array}{rrrr}  
 1 & i & i & -1 \\ 1 & -i & -i & -1 \\ 1 & i & -i & 1 \\ 1 & -i & i & 1 
 \end{array}\right), & \\
 {\cal B}_3=\frac{1}{2}\left(\begin{array}{rrrr}  
 1 & 1 & -i & i \\ 1 & -1 & i & i \\ 1 & 1 & i & -i \\ 1 & -1 & -i & -i 
 \end{array}\right), &&
 {\cal B}_4=\frac{1}{2}\left(\begin{array}{rrrr}  
 1 & -i & 1 & i \\ 1 & i & -1 & i \\ 1 & i & 1 & -i \\ 1 & -i & -1 & -i 
 \end{array}\right). &
\end{xalignat*}
Note that, in this case, the mutually unbiasedness condition is equivalent to the 
condition that ${\cal B}_i\,{\cal B}_i^\dagger=\one_4$, for every $0 \leq i \leq 4$, 
and each entry of ${\cal B}_i\,{\cal B}_j^\dagger$, for $0 \leq i < j \leq 4$, has 
absolute value equal to $\frac{1}{2}$.

\subsection{Construction for $d=p^m$}

By Theorem~\ref{construction-theo}, to construct $p^m+1$ mutually unbiased bases in
$\cc^{p^m}$, we only need to find $m$ symmetric nonsingular matrices
$B_1,\ldots,B_m\in\mm_m(\cc)$
such that the matrix $\sum_{j=1}^m b_jB_j$ is also nonsingular, for every 
{\em nonzero}\/ vector $(b_1,\ldots,b_m)\in{\ff_p}^m$. Because if this condition 
satisfied then the $p^m$ matrices
\[ \sum_{j=1}^m a_jB_j, \qquad (a_1,\ldots,a_m)\in{\ff_p}^m, \]
satisfy the condition of Theorem~\ref{construction-theo}.

\vspace{4mm}
\noindent
{\bf Example $d=8$.} 
The following eight $3\times3$ matrices
determine a set 9 mutually unbiased bases on $\cc^8$. Let $A_1=\bsym{0}_3$ (the zero
matrix), $A_2=\one_3$, and
\begin{xalignat*}{3}
 A_3=\begin{pmatrix} 0 & 1 & 0 \\ 1 & 1 & 1 \\ 0 & 1 & 1 \end{pmatrix} & & 
 A_4=\begin{pmatrix} 0 & 0 & 1 \\ 0 & 1 & 1 \\ 1 & 1 & 0 \end{pmatrix} & & 
 A_5=\begin{pmatrix} 1 & 1 & 0 \\ 1 & 0 & 1 \\ 0 & 1 & 0 \end{pmatrix} &  \\
 A_6=\begin{pmatrix} 1 & 0 & 1 \\ 0 & 0 & 1 \\ 1 & 1 & 1 \end{pmatrix} & & 
 A_7=\begin{pmatrix} 0 & 1 & 1 \\ 1 & 0 & 0 \\ 1 & 0 & 1 \end{pmatrix} & & 
 A_8=\begin{pmatrix} 1 & 1 & 1 \\ 1 & 1 & 0 \\ 1 & 0 & 0 \end{pmatrix} &
\end{xalignat*}
Note that these matrices are of the following general form:
\[ a_1\begin{pmatrix} 1 & 0 & 0 \\ 0 & 1 & 0 \\ 0 & 0 & 1 \end{pmatrix} +
   a_2\begin{pmatrix} 0 & 1 & 0 \\ 1 & 1 & 1 \\ 0 & 1 & 1 \end{pmatrix} +
   a_3\begin{pmatrix} 0 & 0 & 1 \\ 0 & 1 & 1 \\ 1 & 1 & 0 \end{pmatrix} ,
   \qquad a_1,a_2,a_3\in\ff_2. \]
%\[ \begin{pmatrix}
%   a & b & c \\ b & a+b+c & b+c \\ c & b+c & a+b
%   \end{pmatrix} . \]

\vspace{4mm}
Wootters and Fields \cite{wootters89} have found the following general construction for
the matrices $B_1,\ldots,B_m$. Let $\gamma_1,\ldots,\gamma_m$ be a basis of 
$\ff_{p^m}$ as a vector space over $\ff_p$. Then any element $\gamma_i\gamma_j\in\ff_{p^m}$
can be written uniquely as
\[ \gamma_i\gamma_j = \sum_{\ell=1}^m b_{ij}^\ell \gamma_\ell . \]
Then $B_\ell=\left(b_{ij}^\ell\right)$; i.e., the $(i,j)^{\mathrm{th}}$ entry of $B_\ell$
is $b_{ij}^\ell$.

\subsubsection{A set of MUB for the case $d=p^2$}

We would like to mention here that for the case $d=p^2$, there is a more explicit
construction. We find $p^2$ matrices $A_1,\ldots,A_{p^2}$ over $\ff_p$ which satisfy 
the conditions of Theorem~\ref{construction-theo}. For this purpose, we let
\[ A_j=\begin{pmatrix} a_j & b_j \\ b_j & sa_j+tb_j \end{pmatrix} , 
   \qquad a_j,b_j\in\ff_p, \]
where $s,t\in\ff_p$ are two constants which their value need to be determined.
By construction, the matrix $A_j$ is symmetric, so we have to choose the values of
the parameters $s$ and $t$ such that $\det(A_j-A_k)\neq0$, for every 
$1\leq j<k\leq p^2$. Let $\alpha=a_j-a_k$ and  $\beta=b_j-b_k$. Then
$(\alpha,\beta) \neq (0,0)$, and we have
\[ \det(A_j-A_k)=D(\alpha,\beta) = \begin{vmatrix} \alpha & \beta \\ 
                                   \beta & s\alpha+t\beta \end{vmatrix}
          = s \alpha^2+t\alpha\beta-\beta^2 . \]  
If $\alpha=0$, then $D(\alpha,\beta) = -\beta^2 \neq 0$. Suppose now that 
$\alpha \neq 0$, and let $\beta / \alpha = \gamma$. Then
\[ D(\alpha,\beta) = -\alpha^2 ( \gamma^2-t\gamma-s).  \]
Thus $D(\alpha,\beta) \neq 0$ if the quadratic polynomial $\gamma^2-t\gamma-s$ is
irreducible over $\ff_p$. Since for every prime $p$ there is at least one irreducible
quadratic polynomial over $\ff_p$, it is possible to choose the parameters $s,t\in\ff_p$
such that $D(\alpha,\beta) \neq 0$, for every $\alpha,\beta\in\ff_p$. 

%Therefore, we have the following result.
%\begin{theo}
%For any prime $p$, there is a set of $p^2+1$ mutually unbiased bases in $\cc^{p^2}$.
%\end{theo}

\vspace{4mm}
\noindent
{\bf Example $d=4$.} The four matrices (\ref{d=4-example}) are obtained from
the irreducible polynomial $x^2+x+1$ over $\ff_2$. Therefore, all those matrices 
are of the following form
\[ \begin{pmatrix} a & b \\ a & a+b \end{pmatrix}, \qquad a,b\in\ff_2. \]

\vspace{4mm}
\noindent
{\bf Example $d=9$.} The polynomial $x^2+x+2$ is irreducible over $\ff_3$. Therefore, 
the matrices $A_j$ are of the general form of
\[ \begin{pmatrix}  a & b \\ b & a+2b \end{pmatrix} . \]
So the nine matrices are 
\begin{xalignat*}{3}
 \begin{pmatrix} 0 & 0 \\ 0 & 0 \end{pmatrix}, & & 
 \begin{pmatrix} 1 & 0 \\ 0 & 1 \end{pmatrix}, & & 
 \begin{pmatrix} 2 & 0 \\ 0 & 2 \end{pmatrix}, &  \\
 \begin{pmatrix} 0 & 1 \\ 1 & 2 \end{pmatrix}, & & 
 \begin{pmatrix} 1 & 1 \\ 1 & 0 \end{pmatrix}, & &
 \begin{pmatrix} 0 & 2 \\ 2 & 1 \end{pmatrix}, &  \\
 \begin{pmatrix} 1 & 2 \\ 2 & 2 \end{pmatrix}, & & 
 \begin{pmatrix} 2 & 1 \\ 1 & 1 \end{pmatrix}, & & 
 \begin{pmatrix} 2 & 2 \\ 2 & 0 \end{pmatrix}. &  
\end{xalignat*}

\section{Conclusion}

In this paper we partially solved the problem of existence of sets of MUB in
composite dimensions. We formulated an interesting connection between
maximal commuting basis of orthogonal unitary matrices and sets of MUB. We
obtained the necessary condition for the existence of sets of MUB in any dimension.
Using these we proved the existence of sets of MUB for dimensions which are prime
power. We provided a sharp upper bound on the size of any MUB
for any dimension. We expressed the sets of MUB observables as tensor products of
Pauli matrices. However we could not apply this method when the dimension $d$ is a 
product of different primes instead of being a 
prime power (the simplest case that belongs to this category is when $d=6$)
because if we do so the convenient properties of the case $d=p^m$ no longer remain
valid. For instance Theorem~\ref{construction-theo} does not hold in this case. 

A useful application of our result is in secure key distribution using
higher dimensional quantum systems. Specifically we note that the protocol
suggested by Bechmann--Pasquinucci and Tittel \cite{bechmann00b} using  four dimensional
quantum system will become more efficient if all the five mutually unbiased bases are 
used in the protocol instead of only two as suggested by the authors.

Note added: After we submitted our paper for this journal and posted it on the 
Los Alamos quant--ph web site, a related paper \cite{lawrence} was posted on
that e--print server. In that paper, 
with an approach similar to that introduced by us in this paper, in the case of $d=2^m$,
the authors discuss the relationship between MUB and the commuting bases of unitary 
matrices, similar to what we have presented in this paper.

\end{document}